\documentclass{kluwer}    

\usepackage{graphicx}

\begin{document}

\begin{article}

\begin{opening}         

\title{Evolution of Galaxy morphologies in Clusters\thanks{Based on observations 
            taken at the NOT and 1.5-Danish telescopes.}} 
\author{Giovanni \surname{Fasano}$^{(1)}$, Bianca \surname{Poggianti}$^{(1)}$,
Warrick \surname{Couch}$^{(2)}$, Daniela \surname{Bettoni}$^{(1)}$, 
Per \surname{Kj\ae rgaard}$^{(3)}$, Mariano \surname{Moles}$^{(4)}$} 
\runningauthor{Giovanni Fasano et al.}
\runningtitle{Evolution of Galaxy morphologies in Clusters}
\institute{$^{(1)}$Osservatorio Astronomico di Padova (IT)\\ 
$^{(2)}$University of New South Wales (AU)\\ $^{(3)}$Copenhagen University
Observatory (DK)\\ $^{(4)}$Instituto de Matematicas y Fisica Fundamental (ES)}
\date{August 15, 2000}

\begin{abstract}

We have studied the evolution of galaxian morphologies from
ground-based, good-seeing images of 9 clusters at z=0.09-0.25. The
comparison of our data with those relative to higher redshift clusters
(\citeauthor{Dress97} 1997, D97) allowed us to trace for the first
time the evolution of the morphological mix at a look-back time of 2-4
Gyr, finding a dependence of the observed evolutionary trends on the
cluster properties.

\end{abstract}

\keywords{Galaxies, Clusters of galaxies}

\end{opening}           

\section{Background}  

Over the past five years, thanks to the high spatial resolution
imaging achieved with the HST, it has been established that the
morphological properties of galaxies in rich clusters at intermediate
redshift differ dramatically from those in nearby clusters. The most
obvious difference is the overabundance of spirals in the cluster
cores at $z=0.3-0.5$. The second evidence for morphological evolution
in clusters was uncovered only from post-refurbishment data, mainly
thanks to the so called MORPHS collaboration (D97): coupled to the
increase in the spiral fraction, the S0 galaxies at intermediate
redshifts are found to be proportionately less abundant than in nearby
clusters, while the fraction of ellipticals is already as large or
larger.

Another proof of the changes occurring in clusters is the observed
evolution of the morphology-density (MD) relation -- the correlation
between galaxy morphology and local projected density of galaxies -- that
\citeauthor{Dress80} (\citeyear{Dress80}, D80) found in all types of
clusters at low redshift, whereby the elliptical fraction increases
and the spiral fraction decreases with increasing local density. A MD
relation qualitatively similar to that found by D80 was discovered by
the MORPHS to be present in regular clusters and absent in irregular
ones at $z\sim 0.5$.
Overall, the available data seem to require a strong morphological
evolution in clusters between $z=0.4$ and $z=0$. Still, it is worth
keeping in mind that these conclusions are based on a small sample of
distant clusters and on the comparison of a limited redshift range
around $z\sim 0.4$ with the present-day cluster populations.
The goal of the present work is to fill in the observational
gap between the distant clusters observed with {\it HST} and the
nearby clusters, and hence trace, for the first time, the evolution of
the morphological mix at a look-back time of $2-4$ Gyr.

\section{Observations and morphological classification}

The data we used are part of a different project, for which 25
clusters spanning the redshift range $0.03-0.25$ have been
observed. The observations were collected at the NOT and 1.5~Danish
telescopes. To be consistent with previous morphological studies,
among the 25 clusters observed, we have selected 9 clusters for which
an acceptable coverage of the central 1~$\rm Mpc^2$ has been
imaged. As in D97, the analysis of the morphological types has been
done for galaxies down to a visual absolute magnitude $M_V \sim
-20.0$.
In order to improve the morphological type estimates, besides the visual
inspection of the images, we have used the luminosity and geometrical 
profiles obtained with the automatic surface photometry tool GASPHOT
\cite{Pignatelli}.

We have devised four different 'blind' tests to check the reliability
of our morphological classifications both in an absolute sense and
relative to the MORPHS scheme. In particular, the absolute accuracy of our
classifications has been checked in two ways: {\sl (i)} by means of `toy'
galaxies with bulge/disk luminosity and size ratios typical of E, S0
and Sp galaxies; {\sl (i)} by producing redshifted versions of several
galaxies of different morphological types belonging to the nearby
galaxy imaging collection of \citeauthor{Frei} (\citeyear{Frei}). In
both tests the proper values of the observing parameters have been
used to mimic our images and the resulting galaxies have been
classified following the procedure used for our cluster galaxies.  

\section{Results}

\begin{figure}[t] 
\centerline{\includegraphics[width=8cm]{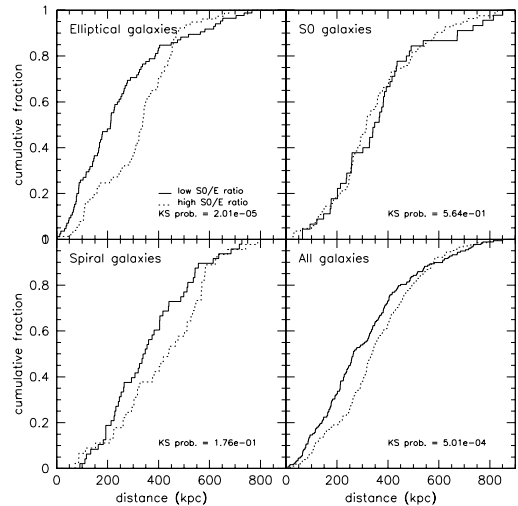}}
\caption{Kolmogorov--Smirnov test applied to the overlapped radial
distributions of galaxies of different types for
low-S0/E and high-S0/E clusters.}
\label{Fig1}
\end{figure}

The clusters appear to be grouped in two different families, according
to their S0/E ratios: a low S0/E family with ratio $\sim$0.8, and a
high S0/E family with ratio $\ge$1.6. We have found that the only structural
difference between the low- and the high-S0/E clusters is the
presence/absence of a high concentration of elliptical galaxies in the
cluster centre ($HEC$/$LEC$ clusters, respectively). A quantitative
illustration of the difference in the galaxy spatial distribution
between the low- and the high-S0/E clusters is given by the KS test
shown in Fig.~\ref{Fig1}. 

\begin{figure}[t] 
\centerline{\includegraphics[width=10cm]{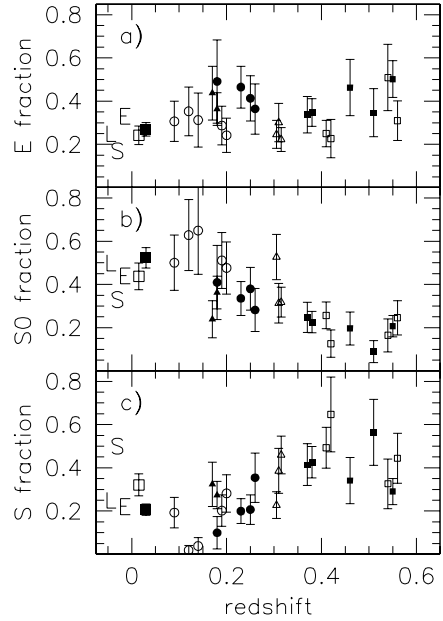}}
\caption{
Morphological fractions as a function of redshift. $HEC$ and $LEC$ clusters
are displayed as solid and open symbols, respectively. The
values from our sample are indicated by circles, whereas those
from the MORPHS and C98+ samples are indicated with squares
and triangles, respectively. 
}
\label{Fig2}
\end{figure}

We have investigated the evolution of the galactic morphologies by
comparing our results with other studies at lower and higher redshift.
At higher redshifts, we have considered the MORPHS sample plus five
additional clusters in the range $z=0.2-0.3$, which we call the C98+
sample. The morphological fractions of all the clusters as a function
of redshift are presented in Fig.~\ref{Fig2}. At low redshift, we
refer to D80 as local benchmark for the $HEC$ and $LEC$ clusters 
(solid and open large squares in Fig.~\ref{Fig2}, respectively). We
have also considered the values quoted by \citeauthor{Oemler} 
(\citeyear{Oemler}) for different
cluster types: spiral/elliptical/S0~-rich (S/E/L in Fig.~\ref{Fig2},
respectively).
In spite of the large error bars, it is clear from this figure that there
are systematic trends with the redshift: moving towards lower redshift
the spiral fraction declines and the S0 fraction rises.  The
morphological fractions in our clusters are intermediate between the
high and the low redshift values and seem to trace a continuous change
of the abundances.  In contrast, the elliptical fraction shows no
particular trend with redshift, but rather a large scatter from
cluster to cluster at any epoch.

We have also analyzed the MD relation of our clusters. We found
that, as in the higher redshift clusters, a morphology-density
relation is present in highly concentrated clusters and absent in the
low concentration ones. This suggests that the morphology-density
relation in low concentration clusters was established in the last 1-2
Gyr.

\end{article}
\end{document}